\documentclass[12pt]{article}
\usepackage{psfig}
\usepackage{sao1}

\begin{document}

\title{Spectroscopy of ``Big Trio'' Objects Using the ``Scorpio''
Spectrograph of the 6-m Telescope of the Special Astrophysical Observatory
\footnote{\small
\noindent
\mbox{ISSN\,1063-7729, Astronomy\,Reports, 2010, {\bf 54}, Iss.8, 675-695.}\\
Original Russian Text (C) Astronomicheskij Zhurnal, 2010,{\bf 87}, No.8,
739-759.
}
}

\author{
Yu.~N.~Parijskij$^1$
\and A.~I.~Kopylov$^1$
\and A.~V.~Temirova$^2$
\and N.~S.~Soboleva$^2$
\and O.~P.~Zhelenkova$^1$
\and O~V.~Verkhodanov$^1$
\and W.~M.~Goss$^3$
\and T.~A.~Fatkhullin$^1$
}
\institute{
Special Astrophysical Observatory, Russian Academy of Sciences,
  Nizhnij Arkhyz, Karachaj-Cherkessian Republic, 357147 Russia
\and St. Petersburg Branch of the Special Astrophysical Observatory,
  Russian Academy of Sciences,
   Pulkovskoe sh. 65, St. Petersburg, 196140 Russia
\and National Radio Astronomy Observatory,
  P. O. Box O, 1003 Lopezville Road, Socorro, NM 87801-0387, USA
}
\date{February 1, 2010}{February 5, 2010}
\maketitle
\begin{abstract}
We present the results of spectroscopy of 71 objects with steep
and ultra-steep spectra ($\alpha<-0.9$, $S\propto\nu^\alpha$)
from the ``Big Trio'' (RATAN-600--VLA--BTA) project, performed
with the ``Scorpio''
spectrograph on the 6-m telescope of the Special Astrophysical Observatory
(Russian Academy of
Sciences). Redshifts were determined for these objects. We also present
several other parameters of the
sources, such as their Rmagnitudes, maximum radio sizes in seconds of arc,
flux densities at 500, 1425, and
3940 MHz, radio luminosities at 500 and 3940 MHz, and morphology.
Of the total number of radio galaxies
studied, four have redshifts $1<z<2$, three have $2<z<3$, one has
$3<z<4$,and one has $z=4.51$.
Thirteen sources have redshifts $0.7<z<1$ and 15 have $0.2<z<0.7$.
Of all the quasars studied, five have
redshifts $0.7<z<1$, seven have $1<z<2$, four have $2<z<3$, and one has
$z=3.57$. We did not detect any spectral lines for 17 objects. \\
{\bf DOI:} 10.1134/S1063772910080019
\end{abstract}

\section*{1. INTRODUCTION}
The ``Big Trio''5 project, which was initiated in
1991-1992 [1] and is based on three large
instruments --- the RATAN-600, the VLA, and the
6-m optical telescope of the Special Astrophysical
Observatory (SAO) -- is aimed at searching for and
studying very distant radio galaxies. (For a detailed
description of the project, see the book by Verkhodanov and Pariiskii [2].)
The project was carried out
in several stages. The first stage -- a search survey
with the RATAN-600 radio telescope in 1980-1981
(the ``Kholod'' experiment [3, 4]) -- resulted in the RC
catalog [5, 6] at 3940 MHz, which also made use of 
the Texas catalog at 365 MHz
\footnote{Kindly provided by Prof. J. Douglas prior to its publication.}
to obtain a sample
the Texas catalog at 365 MHz1 to obtain a sample
of objects with steep and ultra-steep spectra. (The 
steepness of the radio spectrum was already known 
to be useful for selecting very distant objects [7, 8].) 

The next stage was to study the structure of the 
sample objects with the VLA at 1425 and 4885 MHz.
This made it possible to identify FRII sources -- the
most energetic radio galaxies, associated with gE
giant elliptical galaxies - and to improve their coordinates.

The third stage was
optical identification of the
radio sources: for bright sources, from the Palomar
prints and, for fainter ones, from R-band observations with the 6-m
telescope of the SAO [9-11]. We
studied the optical structures of some objects with
a resolution of $<$1'' using the NOT (Nordic Optical
Telescope, Canary Islands) [12].
 
To derive the spectral energy distributions of the 
host galaxies based on evolutionary models, the 
fourth stage included BVRI-photometry with the
6-meter SAO telescope [13, 14]. These data were used
to obtain photometric redshifts and estimate the ages 
of the host galaxies.

In the last stage, we determined spectroscopic 
redshifts, first with the multi-pupil spectrograph [15] 
and then with the ``Scorpio'' spectrograph (mounted
on the 6-m telescope of the SAO) [16]. 

The results for all previous stages were published
earlier [9--27]. In this paper, we present the results
of the last stage: spectoscopic redshifts derived from
observations with the ``Scorpio'' spectrograph (on the
6-m telescope of the SAO).

{
\begin{table}
{Table 1. Radio galaxies with $z>0.7$.\\
}\\
\small
\begin{tabular}{|rlccllllll|}
\hline
No. & RCJ name & RA & Dec & $m_R$ & $z_{sp}$ & LAS & $S_{3940}$ & $S_{1400}$ & $S_{500}$  \\
    &          &    &     &       &          &     &     mJy    &    mJy     &     mJy    \\
1   &  2       &  3 &  4  &   5   &     6    &   7 &      8     &   9        &       10   \\
\hline
1 & 0015+0501&00$^h$15$^m$22.8$^s$ & 05$^\circ$01'22.4'' & 23.2 & 0.813 & 20.1'' & 34 & 110 & 295 \\
2 & 0034+0513 & 00 34 06.24  & 05 14 57.6  & 23.3 & 0.962 & 12.1   & 93 & 252 & 605 \\
3 & 0105+0501 & 01 05 34.46  & 05 01 09.8  & 22.8 & 3.138 & 7.6    & 27 &  92 & 285 \\
4 & 0213+0516 & 02 13 36.25  & 05 18 19.2  & 22.1 & 0.935 & 36.8   &143 & 425 & 963 \\
5 & 0225+0506 & 02 25 09.75  & 05 08 37.4  & 22.1 & 0.770 & 0.2    & 82 & 239 & 586 \\
6 & 0311+0507 & 03 11 47.99  & 05 08 04.1  & 22.9 & 4.514 & 2.8    &135 & 537 &1711 \\
7 & 0444+0501 & 04 44 17.93  & 05 01 25.7  & 23.0 & 1.820 & 11.1   & 70 & 206 & 553 \\
8 & 0506+0508 & 05 06 25.00  & 05 08 19.3  & 21.6 & 0.817 & 0.8    & 91 & 221 & 427 \\
9 & 0744+0500 & 07 44 52.63  & 05 00 09.4 &$>$24.5& 2.48  & 10.8   & 28 &  99 & 341 \\
10& 0837+0446 & 08 37 29.37  & 04 44 21.8  & 22.2 & 1.769 & 3.9    & 62 & 164 & 384 \\
11& 0909+0445 & 09 09 51.09  & 04 44 22.9  & 20.6 & 0.753 & 0.1    & 70 & 192 & 472 \\
12& 1322+0449 & 13 22 03.48  & 04 48 50.5  & 20.4 & 0.799 & 1.7    & 46 & 115 & 269 \\
13& 1339+0445 & 13 39 37.85  & 04 55 04.3  & 22.3 & 0.740 & 34.0   & 40 & 126 & 290 \\
14& 1357+0453 & 13 57 37.39  & 04 53 16.1  & 21.3 & 0.864 & 12.4   & 91 & 250 & 584 \\
15& 1503+0456 & 15 03 59.70  & 04 56 50.4  & 22.8 & 0.788 & 4.5    & 64 & 199 & 545 \\
16& 1510+0438 & 15 10 12.67  & 04 39 31.5  & 22.1 & 0.870 & 3.4    & 71 & 154 & 318 \\
17& 1626+0448 & 16 26 50.29  & 04 48 51.3  & 22.9 & 2.656 & 2.4    & 57 & 200 & 564 \\
18& 1638+0450 & 16 38 32.21  & 04 49 56.3  & 22.3 & 1.272 & 1.9    & 177& 436 &1680 \\
19& 2029+0456 & 20 29 43.40  & 04 56 11.3  & 21.7 & 0.789 & 28.9   & 73 & 153 & 364 \\
20& 2224+0513 & 22 24 17.89  & 05 13 47.3  & 21.3 & 0.974 & 36.1   &120 & 328 & 790 \\
21& 2247+0507 & 22 47 15.18  & 05 08 09.0  & 22.1 & 1.055 & 5.6    &120 & 347 & 885 \\
22& 2348+0507 & 23 48 32.01  & 05 07 33.8  & 22.8 & 2.014 & 4.7    &145 & 399 &1151 \\
\hline
\end{tabular}
\vspace*{1ex}\\
\begin{tabular}{|rrrccll|}
\hline
No. &  $\alpha_{3940}$ & $\alpha_{500}$ &  $L_{3940}$ & $L_{500}$ &  Morphology & Note   \\
    &                  &                &  WHz$^{-1}$ & WHz$^{-1}$&             &        \\
1   &       11         &        12      &      13     &       14  &       15    &  16    \\
\hline
1 &  -1.221 & -0.84  & $8.77\times10^{25}$& $5.96\times10^{26}$& T,BC,FRII & abs \\
2 &  -0.817 & -1.031 & $2.63\times10^{26}$& $1.98\times10^{27}$& D,FRII    & abs \\
3 &  -1.154 & -1.142 & $1.66\times10^{27}$& $1.57\times10^{28}$& CL        &  n(Ly$\alpha$) \\
4 &  -0.925 & -0.925 & $4.12\times10^{26}$& $3.21\times10^{27}$& T,C,FRII  & abs \\
5 &  -1.091 &  0.789 & $1.70\times10^{26}$& $1.23\times10^{27}$&   P       & BLRG; B + abs\\
6 &  -1.366 & -1.095 & $2.38\times10^{28}$& $3.19\times10^{29}$& T,BC,FRII &  n(Ly$\alpha$) \\
7 &  -1.098 & -0.911 & $1.02\times10^{27}$& $8.13\times10^{27}$& D,FRII    & n \\
8 &  -0.961 & -0.531 & $1.99\times10^{26}$& $7.24\times10^{26}$& D,FRII    & abs \\
9 &  -1.275 & -1.095 & $7.33\times10^{26}$& $1.02\times10^{28}$& D,FRII    & n(Ly$\alpha$) \\
10&  -0.945 & -0.83  & $7.46\times10^{26}$& $4.39\times10^{27}$& D, FRII   & BLRG \\
11&  -0.927 & -0.927 & $1.26\times10^{26}$& $8.52\times10^{26}$& P         & n  \\
12&  -0.977 &  0.735 & $9.71\times10^{25}$& $4.93\times10^{26}$& T,BC,FRII/I &  abs \\
13&  -1.1   & -0.806 & $7.74\times10^{25}$& $4.71\times10^{26}$&  D,FRII   & n  \\
14&  -0.899 & -0.899 & $2.18\times10^{26}$& $1.39\times10^{27}$& D,FRII    & n  \\
15&  -1.208 & -0.860 & $1.51\times10^{26}$& $1.04\times10^{27}$& CJ        & BLRG;B+abs \\
16&  -0.728 & -0.728 & $1.54\times10^{26}$& $6.91\times10^{26}$& D,FRII    & n+abs \\
17&  -1.176 & -1.047 & $2.22\times10^{27}$& $1.87\times10^{28}$& D,FRII    & n(Ly$\alpha$) \\
18&  -0.875 & -0.875 & $9.53\times10^{26}$& $5.82\times10^{27}$& T,BC,FRII & n \\
19&  -0.78  & -0.78  & $1.34\times10^{26}$& $6.66\times10^{26}$& T,BC,FRII & n \\
20&  -1.032 & -0.796 & $3.65\times10^{26}$& $2.66\times10^{27}$& D,FRII    & n \\
21&  -0.966 & -0.966 & $4.63\times10^{26}$& $3.40\times10^{27}$& D,C?,FRII & BLRG \\
22&  -1.004 & -1.004 & $2.42\times10^{27}$& $1.92\times10^{28}$& D,FRII    & n \\
\hline
\end{tabular}
\end{table}
}

{
\begin{table}
{Table 2. Radio galaxies with $z<0.7$.\\
}\\
\small
\begin{tabular}{|rlccllllll|}
\hline
No. & RCJ name & RA & Dec & $m_R$ & $z_{sp}$ & LAS & $S_{3940}$ & $S_{1400}$ & $S_{500}$   \\
    &          &    &     &       &          &     &     mJy    &    mJy     &     mJy     \\
1   &  2       &  3 &  4  &   5   &     6    &   7 &      8     &   9        &       10    \\
\hline
 1& 0110+0500  &01$^h$10$^m$13.91$^s$ & 04$^\circ$59'57.6'' & 21.3 & 0.633 & 74.4''& 94   &  247 & 566 \\
 2& 0135+0450  &01 35 37.20 &           04 48 33.8 &        18.4 & 0.372 &  7.8  & 101  &  255 & 665  \\
 3& 0209+0501A &02 09 12.56 &           05 00 52.2 &        18.5 & 0.285 &  0.4  & 33.4 &  86  & 242  \\
 4& 0457+0452  &04 57 53.86 &           04 53 53.5 &        19.4 & 0.482 &   67  & 72   &  199 & 461  \\
 5& 0820+0454  &08 20 56.7  &           04 54 16.8 &        19.3 & 0.539 &  2.0  & 165  &  465 & 1206 \\
 6& 0845+0444  &08 45 31.20 &           04 42 54.9 &        21.4 & 0.650 &  7.6  & 156  &  504 & 1164 \\
 7& 0908+0451  &09 08 21.01 &           04 50 58.3 &        19.6 & 0.525 & 34.5  & 111  &  263 & 700  \\
 8& 1011+0502  &10 12 04.58 &           05 06 14.2 &        22.4 & 0.456 &  1.3  & 68   &  210 & 545  \\
 9& 1124+0456  &11 24 37.43 &           04 56 18.7 &        17.3 & 0.284 & 12.0  & 440  &  991 & 2716 \\
10& 1142+0455  &11 42 20.10 &           04 54 56.0 &        21.0 & 0.605 & 18.7  & 107  &  292 & 652  \\
11& 1155+0444  &11 55 19.24 &           04 43 31.3 &        18.6 & 0.289 & 13.0  & 71   &  159 & 353  \\
12& 1235+0435  &12 35 49.52 &           04 32 56.9 &        21.5 & 0.657 &  9.1  & 61   &  153 & 300  \\
13& 1446+0507  &14 46 17.97 &           05 07 41.1 &        19.3 & 0.273 & 68.0  & 163  &  364 & 746  \\
14& 1646+0501  &16 46 53.31 &           05 01 10.0 &        21.2 & 0.690 & 15.7  & 55   &  106 & 313  \\
15& 1722+0442  &17 22 14.06 &           04 43 17.0 &        20.7 & 0.604 & 21.9  & 266  &  768 & 2049 \\
\hline
\end{tabular}
\vspace*{2ex}\\
\begin{tabular}{|rrrccll|}
\hline
No. & $\alpha_{3940}$ & $\alpha_{500}$ &  $L_{3940}$ & $L_{500}$ & Morphology & Note   \\
    &                 &                &  WHz$^{-1}$ & WHz$^{-1}$&            &        \\
1   &      11         &        12      &      13     &       14  &      15    &  16    \\
\hline
 1& -0.994 &-0.744&$1.20\times10^{26}$& $6.46\times10^{26}$& D,FRII    &n+abs \\
 2& -0.955 &-0.88  &$4.02\times10^{25}$& $2.53\times10^{26}$& TB,C,FRII &n+abs \\
 3& -0.96  &-0.96  &$8.20\times10^{24}$& $4.53\times10^{25}$& P         &abs   \\
 4& -1.059 &-0.731 &$5.27\times10^{25}$& $2.97\times10^{26}$& D,FRII/I  &n+abs \\
 5& -0.993 &-0.993 &$1.50\times10^{26}$& $1.09\times10^{27}$& D,FRII    &n     \\
 6& -1.19Â &-0.753 &$2.34\times10^{26}$& $1.34\times10^{27}$& T,C,FRII  &n+abs \\
 7& -0.944 &-0.845 &$9.28\times10^{25}$& $5.64\times10^{26}$& T,C,FRII  &BLRG;B+abs\\
 8& -1.01  &-1.01  &$4.34\times10^{25}$& $3.47\times10^{26}$& CL        &n     \\
 9& -0.914 &-0.849 &$1.00\times10^{26}$& $6.18\times10^{26}$& D,FRII    &n+abs \\
10& -0.874 &-0.874 &$1.18\times10^{26}$& $7.16\times10^{26}$& T,C,FRII  &n+abs \\
11& -0.778 &-0.778 &$1.61\times10^{25}$& $8.05\times10^{25}$& D,FRII    &n+abs \\
12& -0.822 &-0.720 &$7.80\times10^{25}$& $3.64\times10^{26}$& D,FRII    &abs   \\
13& -0.736 &-0.736 &$3.27\times10^{25}$& $1.50\times10^{26}$& T,WC,FRII &n+ab  \\
14& -0.84  &-0.84  &$7.85\times10^{25}$& $4.47\times10^{26}$& D,FRII    &abs   \\
15& -1.061 &-0.918 &$3.18\times10^{26}$& $2.29\times10^{27}$& D,FRII    &n+abs \\
\hline
\end{tabular}
\end{table}
}

{
\begin{table}
\small
{Table 3. Quasars.\\
}\\
\begin{tabular}{|rlccllllll|}
\hline
No. & RCJ name & RA & Dec & $m_R$ & $z_{sp}$ & LAS & $S_{3940}$ & $S_{1400}$ & $S_{500}$ \\
    &          &    &     &       &          &     &     mJy    &    mJy     &     mJy   \\
1   &  2       &  3 &  4  &   5   &     6    &   7 &      8     &   9        &       10  \\
\hline
 1& 0038+0449& 00$^h$38$^m$34.65$^s$ & 04$^\circ$50'50.5''& 2.446 &   3.3'& 21.2 &  94 & 239 & 640 \\
 2& 0042+0504& 00 42 27.14 &           05 05 24.1       & 1.504 &  24.8 & 19.0 &  93 & 231 & 579 \\
 3& 0126+0502& 01 26 16.13 &           05 02 10.3       & 1.008 &  18.0 & 18.1 &  51 & 151 & 402 \\
 4& 0143+0505& 01 43 33.97 &           05 07 58.0       & 2.135 &   7.4 & 20.6 &  52 & 164 & 485 \\
 5& 0226+0512& 02 26 19.81 &           04 46 32.3       & 1.235 &  10.7 & 20.1 &  98 & 242 & 532 \\
 6& 0459+0456& 04 59 04.28 &           04 55 54.4       & 1.189 &  63.8 & 20.9 &  90 & 251 & 564 \\
 7& 1100+0444& 11 00 11.49 &           04 44 01.4       & 0.890 &   0.3 & 19.1 & 239 & 640 &1453 \\
 8& 1154+0431& 11 54 53.50 &           04 24 12.5       & 0.998 &   6.7 & 19.9 & 331 & 854 &1867 \\
 9& 1251+0446& 12 51 29.50 &           04 46 41.7       & 0.96  &   257 & 19.3 & 204 & 514 &1611 \\
10& 1333+0451& 13 33 07.00 &           04 50 48.6       & 1.405 & 129.5 & 17.3 &  21 & 55  & 123 \\
11& 1456+0456& 14 56 25.79 &           04 56 44.8       & 2.13  &   2.2 & 20.0 & 110 & 288 & 727 \\
12& 1740+0502& 17 40 33.96 &           05 02 42.3       & 3.57  &   4.7 & 22.6 &  36 & 110 & 288 \\
13& 2013+0508& 20 13 23.48 &           05 10 30.5       & 0.89  &  10.0 & 21.1 &  53 & 138 & 281 \\
14& 2036+0451& 20 36 56.93 &           04 49 52.7       & 0.716 &  56.0 & 19.0 &  80 & 228 & 563 \\
15& 2144+0513& 21 44 27.18 &           05 11 15.2       & 1.01  &   1.9 & 18.8 &  66 & 203 & 526 \\
16& 2225+0523& 22 25 14.72 &           05 27 09.1       & 2.323 &   2.7 & 17.8 & 309 & 849 &2219 \\
17& 2320+0459& 23 20 44.74 &           04 59 24.9       & 1.39  &  15.2 & 20.4 &  78 & 169 & 423 \\
\hline
\end{tabular}
\vspace*{2ex}\\
\begin{tabular}{|rrrccll|}
\hline
No. & $\alpha_{3940}$ & $\alpha_{500}$ &  $L_{3940}$ & $L_{500}$ &  Morphology & Note   \\
    &                 &                &  WHz$^{-1}$ &   WHz$^{-1}$&           &        \\
1   &      11         &        12      &      13     &       14  &       15    &  16    \\
\hline
 1& -1.08  &-0.776 &$2.68\times10^{27}$ & $1.25\times10^{28}$ & FRII D. & B(Ly$\alpha$)  \\
 2& -0.937 &-0.836 &$7.61\times10^{26}$ & $4.32\times10^{27}$ & D, FRII & B            \\
 3& -1.133 &-0.874 &$1.99\times10^{26}$ & $1.31\times10^{27}$ & D, FRII & B            \\
 4& -1.245 &-0.926 &$1.31\times10^{27}$ & $8.47\times10^{27}$ & D, FRII & B(Ly$\alpha$)  \\
 5& -0.947 &-0.686 &$5.33\times10^{26}$ & $2.34\times10^{27}$ & D, FRII & B            \\
 6& -1.167 &-0.615 &$5.27\times10^{26}$ & $2.14\times10^{27}$ & D, FRII & B            \\
 7& -0.874 &-0.874 &$5.98\times10^{26}$ & $3.63\times10^{27}$ & P       & B            \\
 8& -0.994 &-0.682 &$1.15\times10^{27}$ & $5.22\times10^{27}$ & D, FRII & B            \\
 9& -1.129 &-0.875 &$7.12\times10^{26}$ & $4.74\times10^{27}$ & D, FRII & B            \\
10& -1.09Â &-0.702 &$1.69\times10^{26}$ & $7.07\times10^{26}$ & T,C,FRII & B           \\
11& -0.875 &-0.875 &$1.80\times10^{27}$ & $1.19\times10^{28}$ & D?      & B            \\
12& -1.148 &-0.870 &$2.71\times10^{27}$ & $1.43\times10^{28}$ & CL      & B (Ly$\alpha$) \\
13& -0.809 &-0.809 &$1.27\times10^{26}$ & $6.74\times10^{26}$ & D, FRII & B            \\
14& -1.076 &-0.816 &$1.40\times10^{26}$ & $8.58\times10^{26}$ & T,C,FRII & B           \\
15& -1.066 &-1.066 &$2.47\times10^{26}$ & $1.97\times10^{27}$ & T,C,FRII & B           \\
16& -0.995 &-0.902 &$7.04\times10^{27}$ & $4.50\times10^{28}$ & T,C,FRII & B           \\
17& -0.868 &-0.769 &$5.05\times10^{26}$ & $2.52\times10^{27}$ & D, FRII & B            \\
\hline
\end{tabular}
\end{table}
}

{
\begin{table}
{Table 4. Radio galaxies with no emission lines detected.
\\
}\\
\small
\begin{tabular}{|rlcclll|}
\hline
No. & RCJ name & RA & Dec & $m_R$ & LAS & $S_{3940}$ \\
    &          &    &     &       &     &     mJy    \\
1   &  2       &  3 &  4  &   5   &   6 &     7      \\
\hline
 1 & 0015+0503a& 00 15 11.61 & 05 06 39.8 & 22.0 & 20.6 & 19 \\
 2 & 0250+0512 & 02 50 53.57 & 05 16 12.7&$>$24.5&  1.2 & 68 \\
 3 & 0308+0454 & 03 08 33.95 & 04 54 10.3 & 23.3 &  1.2 & 29 \\
 4 & 0324+0442 & 03 24 07.32 & 04 42 01.8 & 22.4 & 11.8 &120 \\
 5 & 0355+0449 & 03 55 12.72 & 04 40 41   & 24.2 &  2.4 & 62 \\
 6 & 0446+0525 & 04 46 23.17 & 05 40 50.2 & 23.2 &19.17 &107 \\
 7 & 0743+0455 & 07 43 15.59 & 04 55 52.5 & 23.6 & 20.5 & 37 \\
 8 & 0836+0511 & 08 36 48.09 & 05 13 09.0 & 22.6 & 19.6 &113 \\
 9 & 0945+0454 & 09 45 26.81 & 04 54 13.7 & 23.5 &  4.5 & 20 \\
10 & 1051+0449 & 10 51 25.78 & 04 49 43.9 & 22.5 &  1.7 &122 \\
11 & 1148+0455 & 11 48 47.88 & 04 55 24.4 & 23.2 & 43.9 &250 \\
12 & 1152+0449 & 11 52 23.67 & 04 48 14.3 & 22.5 &  7.0 & 46 \\
13 & 1703+0502 & 17 03 29.32 & 05 02 11.6 & 23.5 &  1.8 &210 \\
14 & 1720+0455 & 17 20 04.6  & 04 53 48.8 &(20.6)& <0.5 & 19 \\
15 & 1735+0454 & 17 35 41.76 & 04 55 15.3 & 23.2 &  5.2 & 28 \\
16 & 2219+0458 & 22 19 06.13 & 04 58 45.7 & 23.8 &  4.7 & 50 \\
17 & 2236+0454 & 22 36 51.17 & 04 55 09.2 & 23.6 & 40.2 & 41 \\
\hline
\end{tabular}
\vspace*{2ex}\\
\begin{tabular}{|rllll|}
\hline
No. & $\alpha_{3940}$ & $\alpha_{500}$ & Morphology & Exposure time and note   \\
1   &       8         &        9  &       10    &  11    \\
\hline
 1 & -1.67  &-1.03  & D, FRII          & 2$\times$900s   \\
 2 & -1.26  &-1.26  & D, FRII          & 4$\times$900s   \\
 3 & -0.965 &-0.965 & CL or D, BC,FRII & 4$\times$900s,3$\times$1200s,6$\times$900s\\
 4 & -1.131 &-0.974 & D, FRII/I        & 6$\times$900s,4$\times$900s \\
 5 & -1.494 &-1.296 & D, FRII          & 2$\times$900s,4$\times$900s \\
 6 & -0.983 &-0.983 & D, FRII          & 3$\times$900s   \\
 7 & -1.064 &-1.064 & D, FRII          & 2$\times$900s   \\
 8 & -1.05  &-1.05  & D, CJ, FRII/I    & 4$\times$120s   \\
 9 & -1.01  &-1.01  & D, FRII          & 4$\times$900s   \\
10 & -0.92  &-0.92  & D, FRII          & 2$\times$900s   \\
11 & -1.24  &-0.94  & D, FRII          & 2$\times$900s   \\
12 & -0.762 &-0.762 & D, FRII          & 3$\times$900s   \\
13 & -1.05  &-1.05  & D, FRII          & 3$\times$900s, 2$\times$600,900s, 2$\times$900s\\
14 & -1.01  &-1.41  & P                & 600s; M star by spectrum\\
15 & -0.924 &-0.924 & D, FRII          & 4$\times$900s   \\
16 & -1.227 &-0.861 & D, FRII          & 2$\times$900s   \\
17 & -1.517 &-0.899 & D, FRII          & 2$\times$900s, 4$\times$1200s \\
\hline
\end{tabular}
\end{table}
}

\begin{figure}
\centerline{\psfig{figure=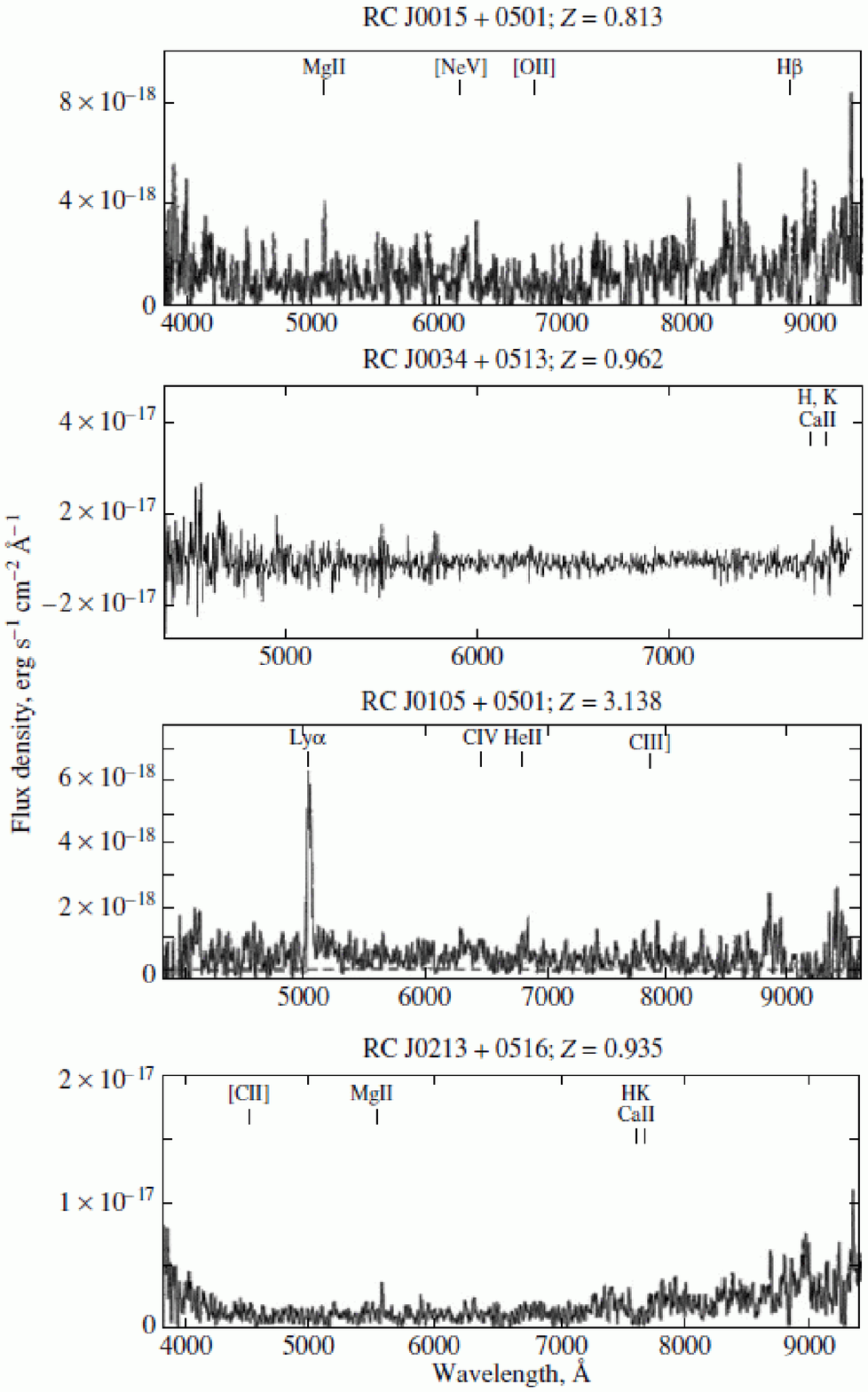,width=17cm}}
\caption{Spectra of radio galaxies with $z>0.7$.}
\end{figure}
\begin{figure}
\centerline{\psfig{figure=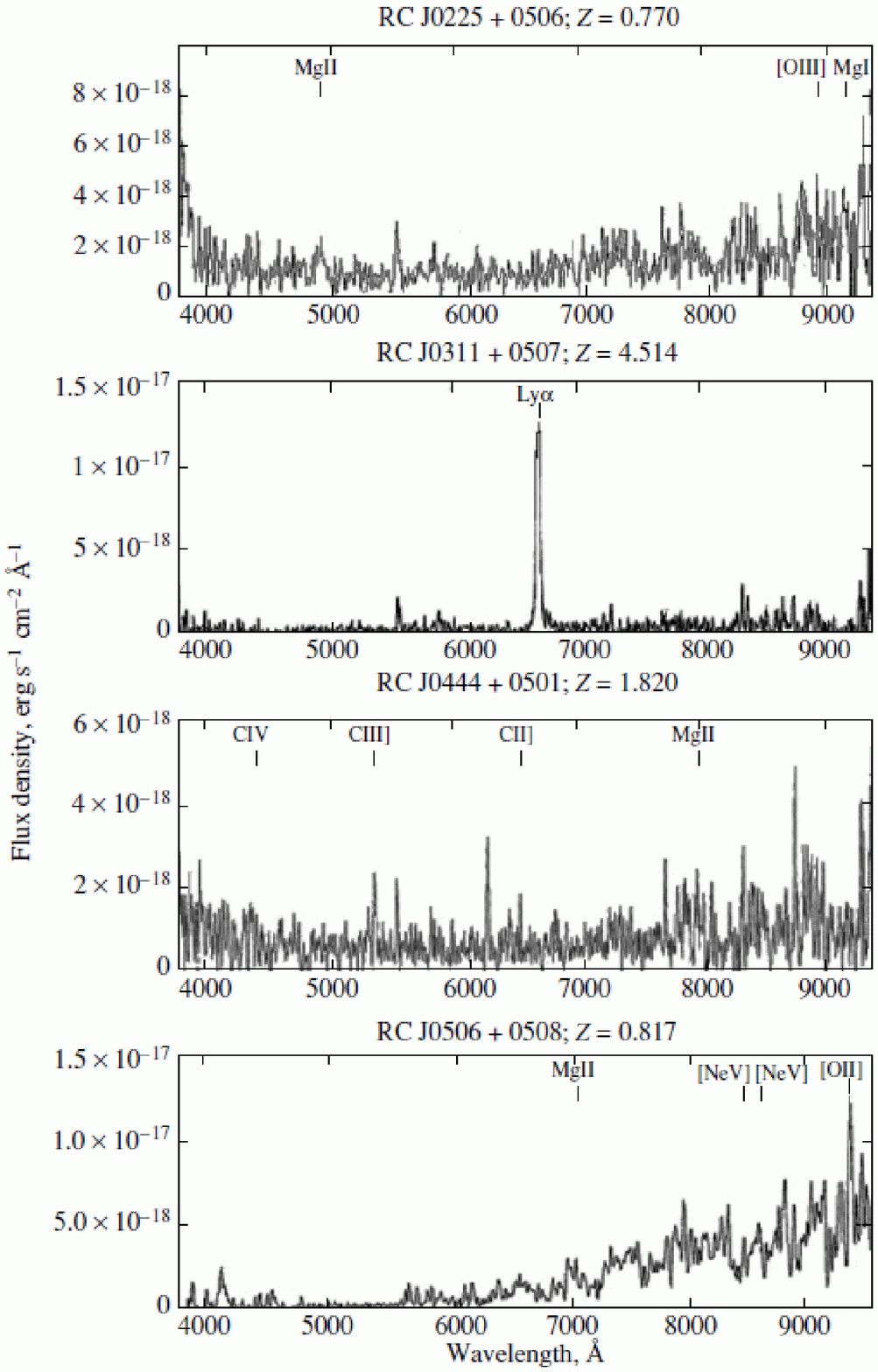,width=17cm}}
\centerline{{Fig. 1. (Continued)}}
\end{figure}
\begin{figure}
\centerline{\psfig{figure=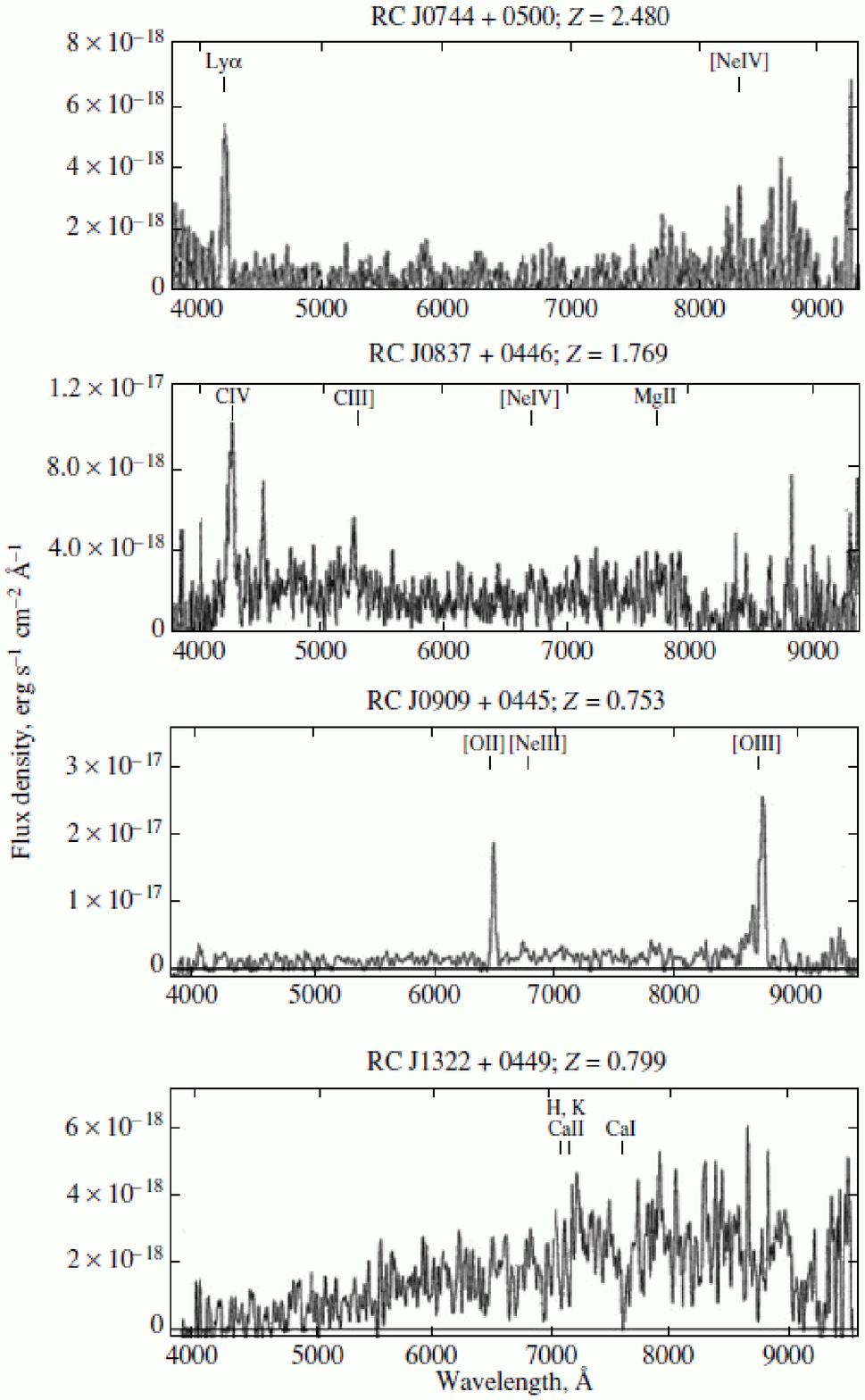,width=17cm}}
\centerline{{Fig. 1. (Continued)}}
\end{figure}
\begin{figure}
\centerline{\psfig{figure=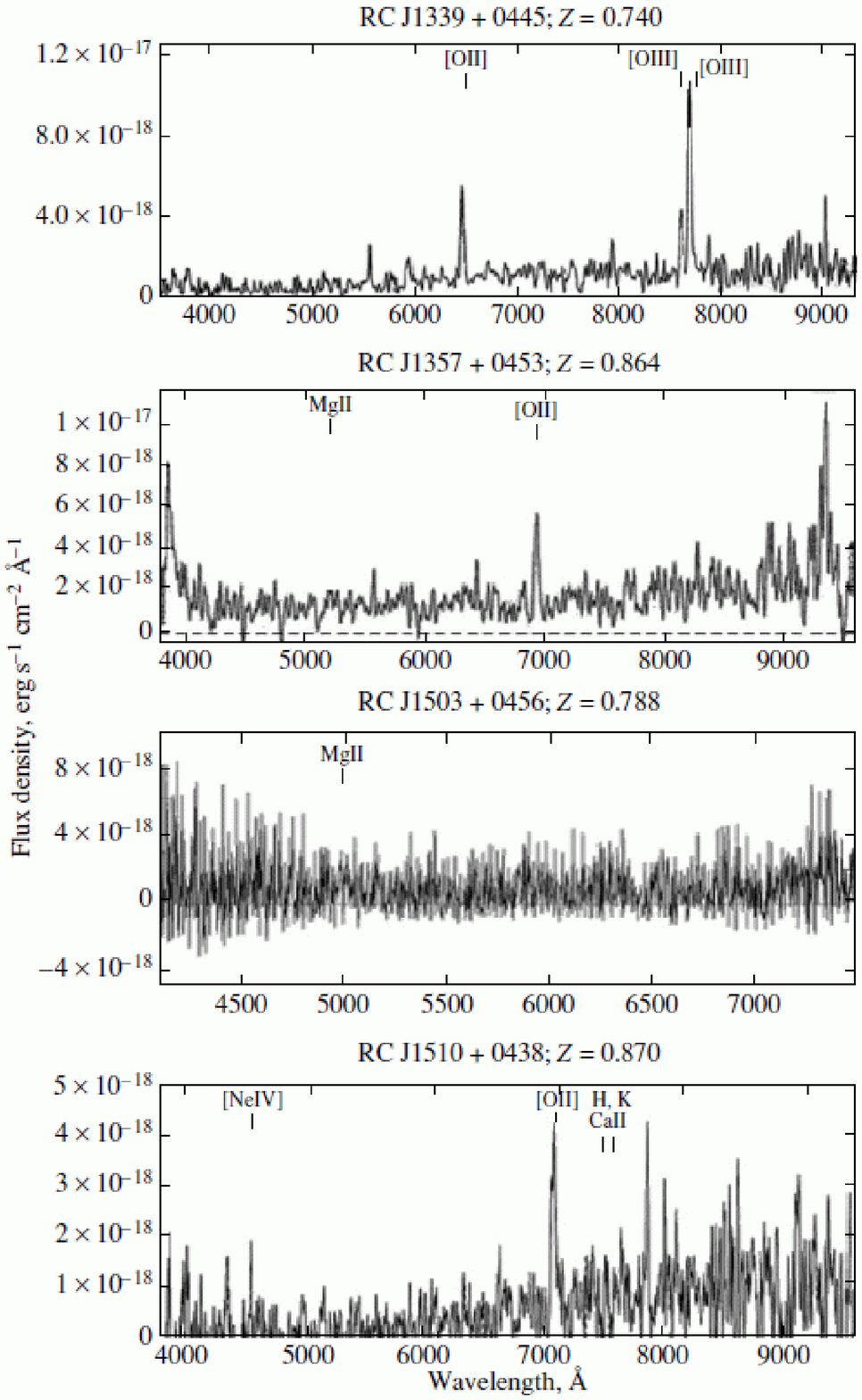,width=17cm}}
\centerline{{Fig. 1. (Continued)}}
\end{figure}
\begin{figure}
\centerline{\psfig{figure=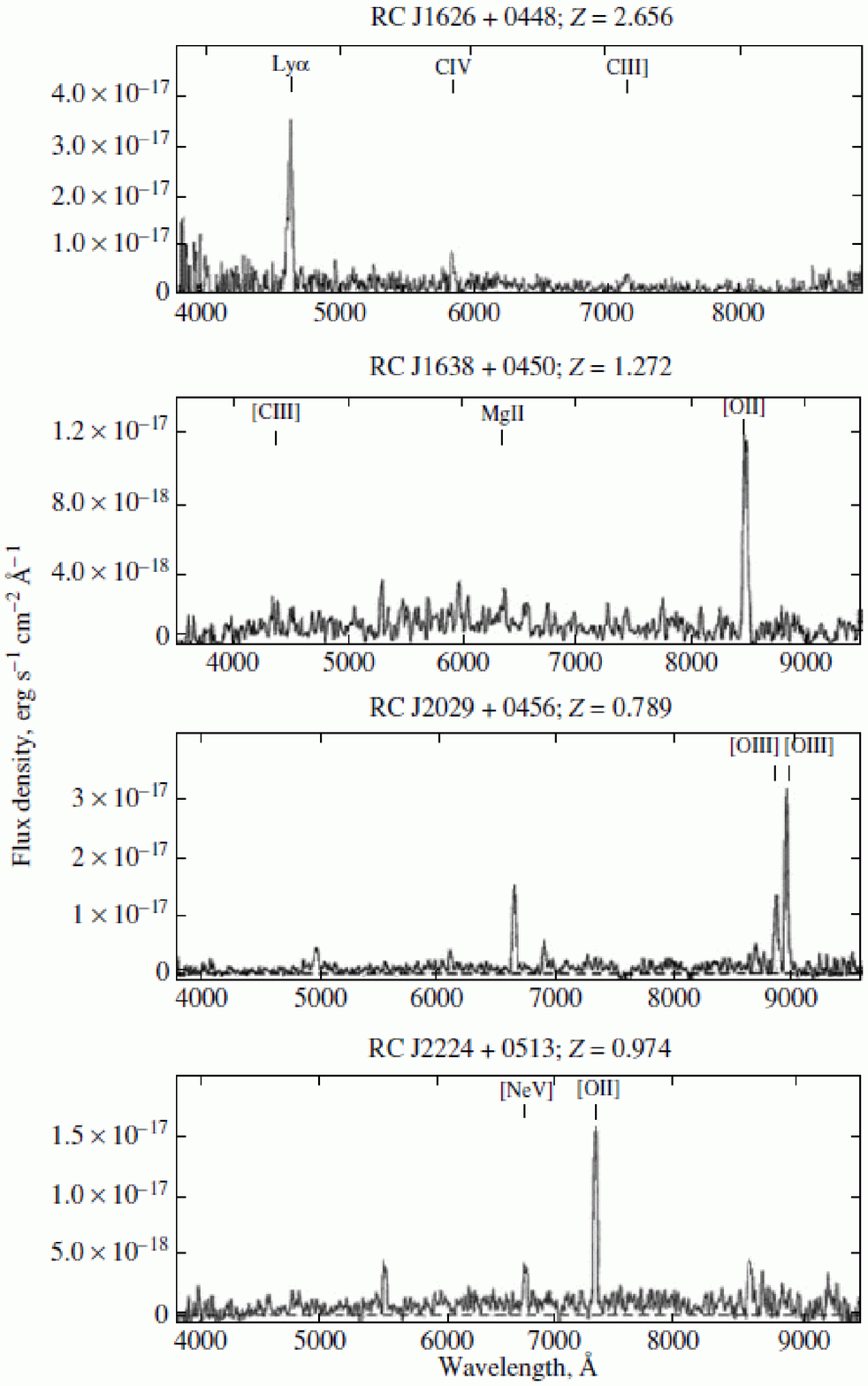,width=17cm}}
\centerline{Fig. 1. (Continued)}
\end{figure}
\begin{figure}
\centerline{\psfig{figure=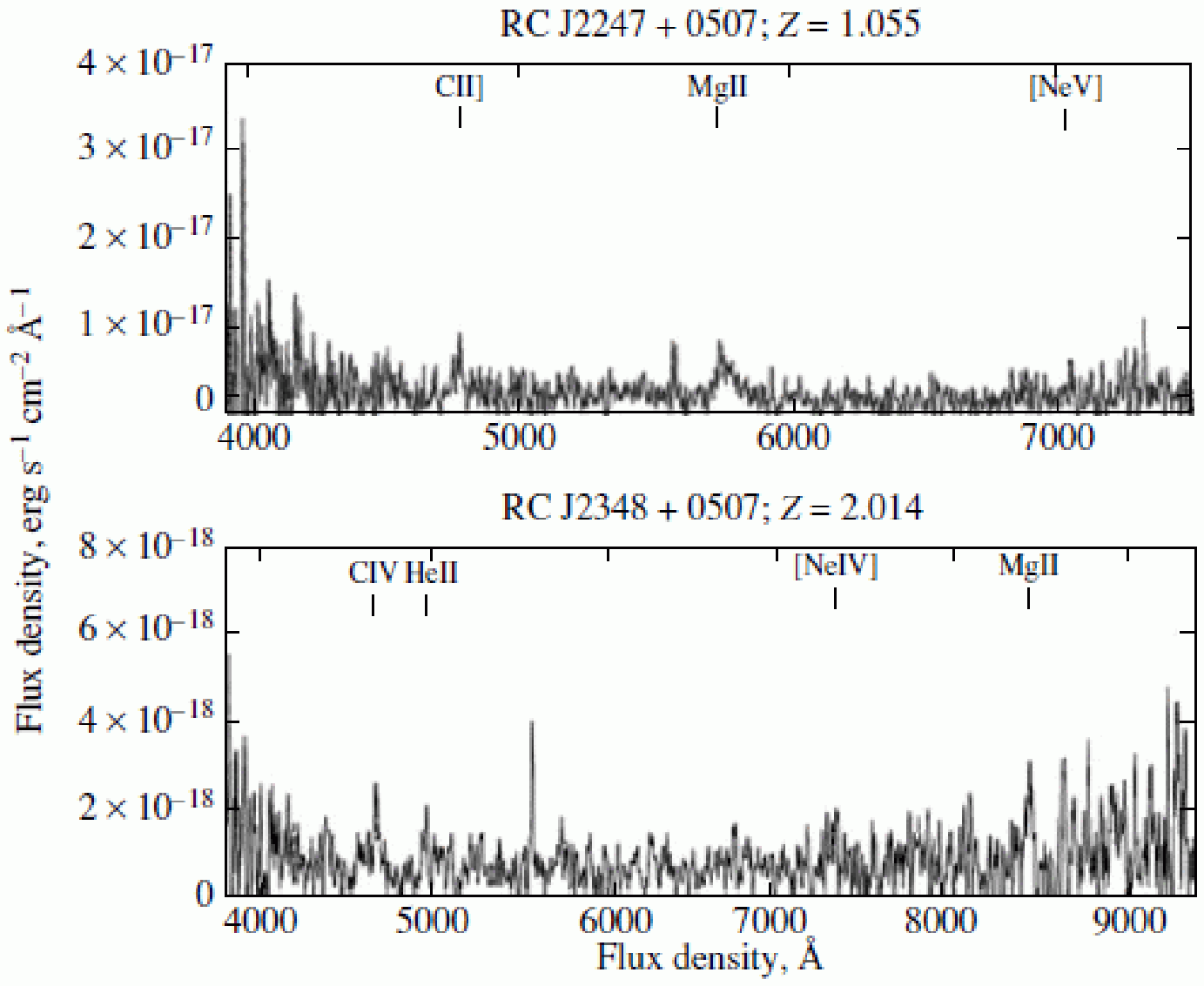,width=17cm}}
\centerline{Fig. 1. (Continued)}
\end{figure}

\section*{2. OBSERVATIONAL RESULTS}

The data obtained from our observations with the 
``Scorpio'' spectrograph mounted on the 6-m SAO
telescope are collected in four tables. In total, we 
studied 71 objects with steep and ultra-steep spectra
 from the RC catalog. These include 17 quasars,
with the other objects being radio galaxies. Tables 1
and 2 present the results for radio galaxies with detected
 emission lines, subdivided according to their 
redshifts. Table 1 contains 22 radio galaxies with 
redshifts z>0.7,and Table 2 -- 15 radio galaxies with
redshifts between 0.2 and 0.7. 

The columns of Tables 1 and 2 contain for each
object:
1 --- a running number;
2 ---- the name in the RC catalog;
3, 4 --- the right ascension and declination of the
object for equinox 2000;
5 --- the R-band magnitude;
6 --- the spectroscopic redshift ($z_{sp}$);
7 --- the largest angular size in arcseconds, LAS;
8, 9, 10 --- the flux densities S, in mJy, at 3940,
1400, and 500 MHz (the last obtained from interpolating the spectrum);
11, 12 --- the spectral indices $\alpha$ ($S\propto\nu^\alpha$)
at 3940 and 500 MHz;
13, 14 --- the luminosities L, W Hz$^{-1}$, at 3940 and 500 MHz;
15 --- the radio source morphology, with `P' denoting
a point-like object, `D' a double source, `T' a triple
source, `C' the presence of a core, `BC' the presence of
a bright core, `WC' the presence of a weak core, `CL'
a core and lobe structure (distinct components), `CJ'
a core and jet structure (the core not distinguishable 
from the extended jet), and FRI/FRII an object of 
Fanaroff--Riley types I/II;
16 --- notes, with n, B and abs indicating the presence
 of narrow emission lines, broad emission lines, 
and absorption lines in the spectrum and BLRG indicating
 a broad-line radio galaxy. 

The columns of Table 3 contain the same information as in Tables 1
and 2 for the 17 detected quasars.
Table 4 presents the results for the 17 radio galaxies 
without detected emission lines. 

Columns 1--5 of Table 4 are analogous to the
corresponding columns in Tables 1--3. The following
columns contain:
6 --- the largest angular size in arcseconds, LAS;
7 --- the flux density, S, in mJy at 3940 MHz;
8, 9 --- the spectral indices $\alpha$ at 3940 and 500 MHz;
10 --- the morphology;
11 --- the exposure time, in seconds, and notes.
Note that the absence of detectable lines for
faint objects could, in some cases, be due to a
higher redshift (z>5) or absorption by dust, or bad
weather conditions (with seeing $\sim$2''--3''). One can
not also exclude an incorrectly placed spectrograph 
slit. For example, such objects in Table 4 include 
RCJ\,0250+512 and RCJ\,0355+0449, whose spectral
indices, morphologies, angular sizes, and ratios of 
the radio to optical luminosity suggest they may be
very distant objects. It would be desirable to obtain
spectroscopy of these objects using telescopes with
higher sensitivities and broader frequency ranges, 
repeat the 6-m SAO observations under very good 
seeing conditions, or obtain observations in the 
K-filter.

Below we present remarks for several of the radio 
sources. 

{\bf RCJ\,0015+0503a} (Table 4). This object has B=23.98, V=23.04,
R=22.26, I=21.40. The observing
 conditions were poor, with seeing of $\sim$3''; $z_{ph}$=0.60$\pm$0.13.

{\bf RCJ\,0034+0513} (Table 1). The redshift was measured from two
absorption lines, at 7700\,\aa (the KCaII 3933.7\,\aa line) and
7800\,\aa(the HCaII 3968.5\,\aa line);
$z_{ph}$=1.03.

{\bf RCJ\,0105+0501} (Table 1). The radio source is
double; it is not clear where the AGN is located. 
An optical object displaying Ly$\alpha$ emission has the
coordinates $\alpha$=01h05m34.091s, $\delta$=+5$^\circ$01'12.47''
(J2000), not coincident with any of the radio components.
 Radio observations with higher sensitivity and 
resolution are desirable, in order to detect the radio 
core. The object's has B=24.1, V=22.5, R=22.8,
I=22.4. 

{\bf RCJ\,0213+0516} (Table 1). The object is a cluster
member. The brightest member of the cluster is a 
radio quasar at redshift $z_{sp}=0.94$ (approximately the
same $z$ as for the FRII radio galaxy), 8''
to the east 
of the radio core. The contribution from the compact 
core is about 3\% of the total flux density of the radio
source. There is a small jet directed away from the 
quasar. Interaction between the quasar and the host 
galaxy cannot be ruled out. (The VLA radio image 
with overlaid 6-m SAO optical image can be found at 
{\tt http://wo.sao.ru/hd/zhe}; similar information is also
available for other objects at this same address.) 

{\bf RCJ\,0311+0507} (Table 1). This object has the
highest redshift in our sample [26, 27]. It has B$>$24.9, V$>$24.8,
R=22.6, I=22.3. This radio source
has now been studied using the ``Merlin'' radio
interferometer (UK) and European VLBI Network 
(EVN). 

{\bf RCJ\,0324+0442} (Table 4). This is an FRII double
radio source, most probably in an empty field (the two 
galaxies near the eastern component are foreground 
objects); $m_R$$>$25. If such objects are detected in the
K-band at the level K=18--19, they enter the range
$z_{sp}$=1.5--2, and have no bright emission lines in the
optical. 

{\bf RCJ\,0506+0508} (Table 1). The redshift was
determined from absorption lines at 7150--7250\,\aa
(though there remain some doubts). 

{\bf RCJ\,0836+0511} (Table 4). This is not an entirely
classical FRII object: the radio map at 8460\,MHz
(VLA) reveals a core, a jet toward the Northeast, a 
hot spot, and a ``tail''; $z_{ph}$=1.12.

{\bf RCJ\,0908+0451} (Table 2). $z_{sp}$=0.542, based on
data from the NASA Extragalactic Database (NED). 

\begin{figure}
\centerline{\psfig{figure=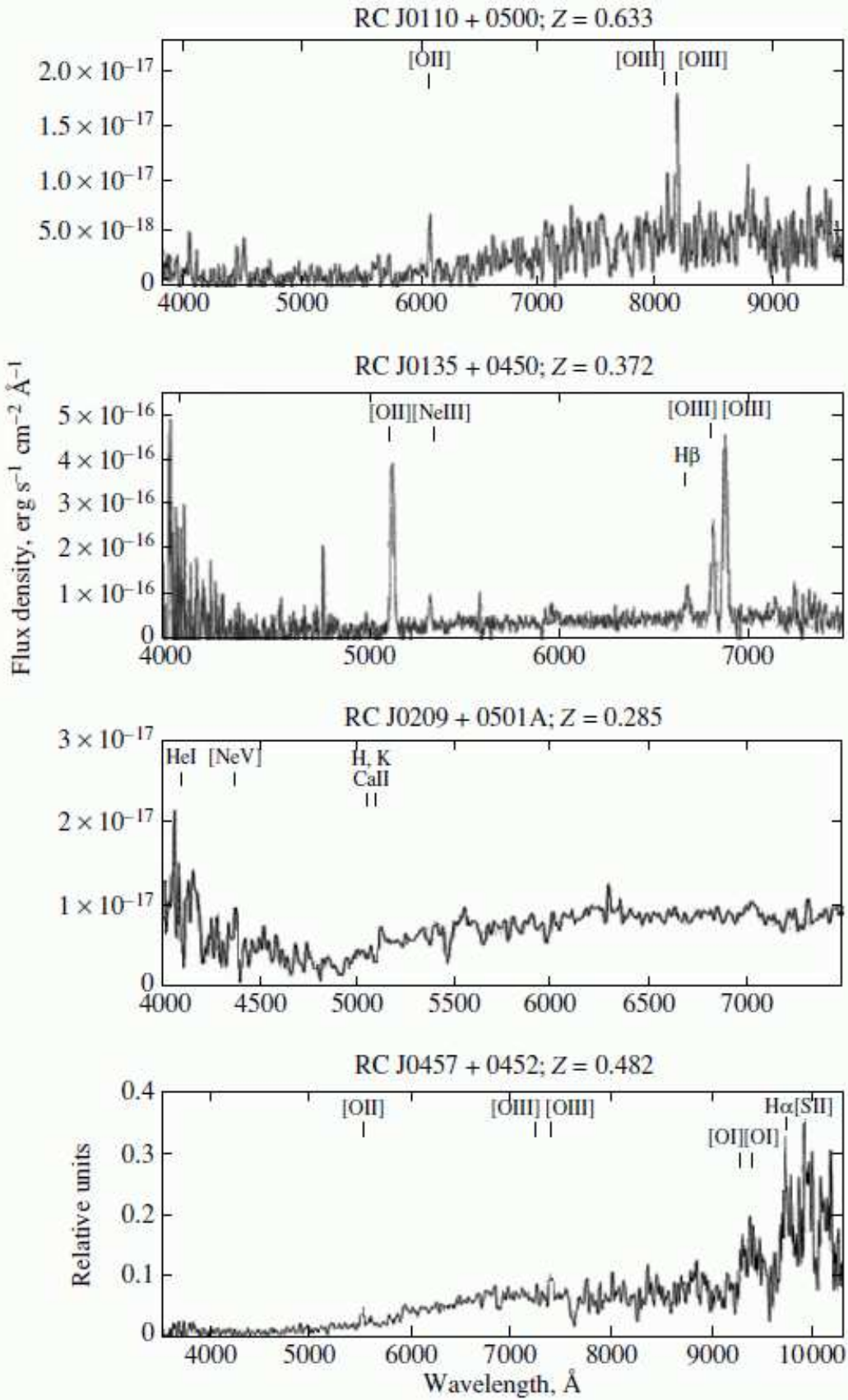,width=17cm}}
\caption{Spectra of radio galaxies with z<0.7.}
\end{figure}
\begin{figure}
\centerline{\psfig{figure=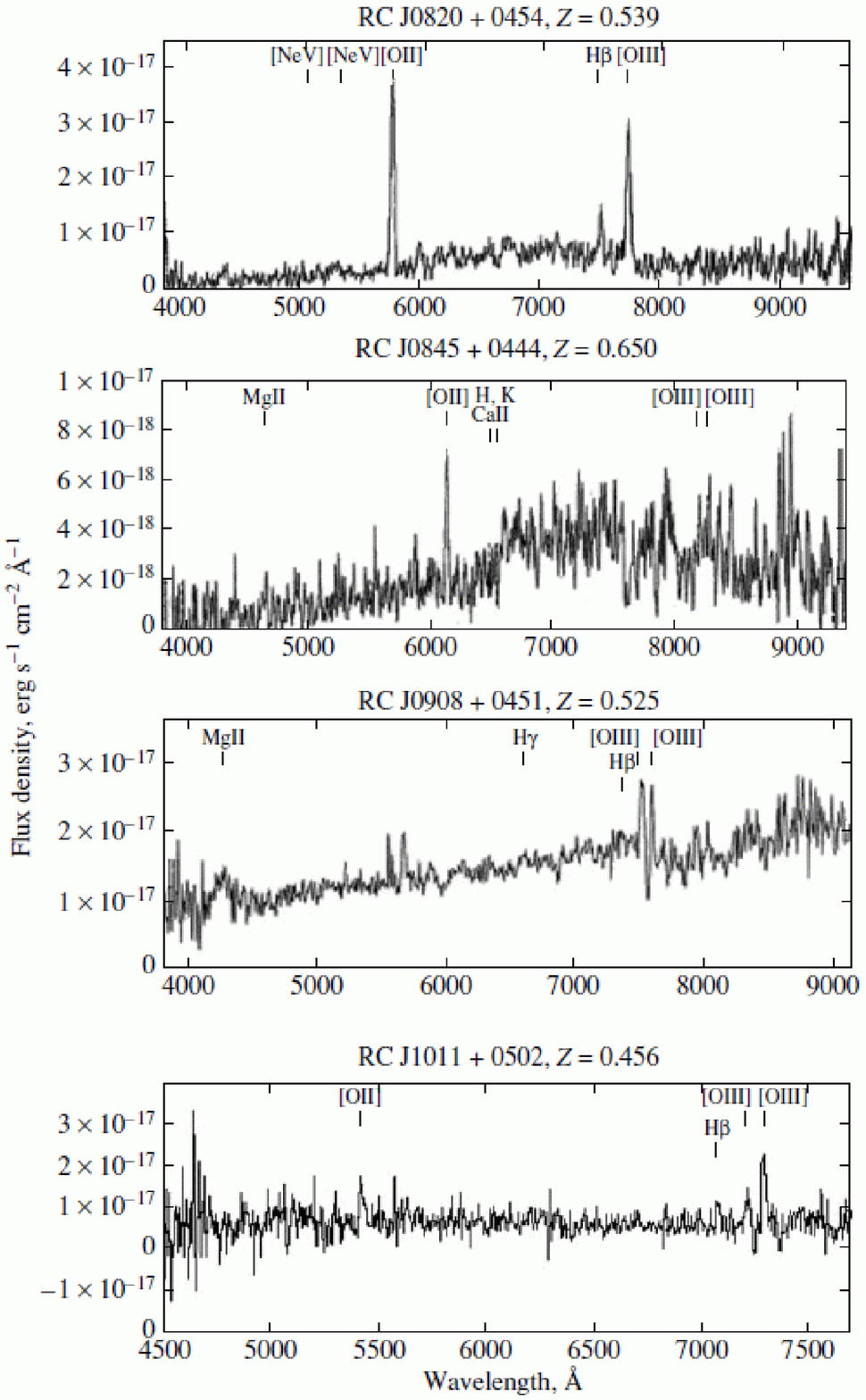,width=17cm}}
\centerline{{Fig. 2. (Continued)}}
\end{figure}
\begin{figure}
\centerline{\psfig{figure=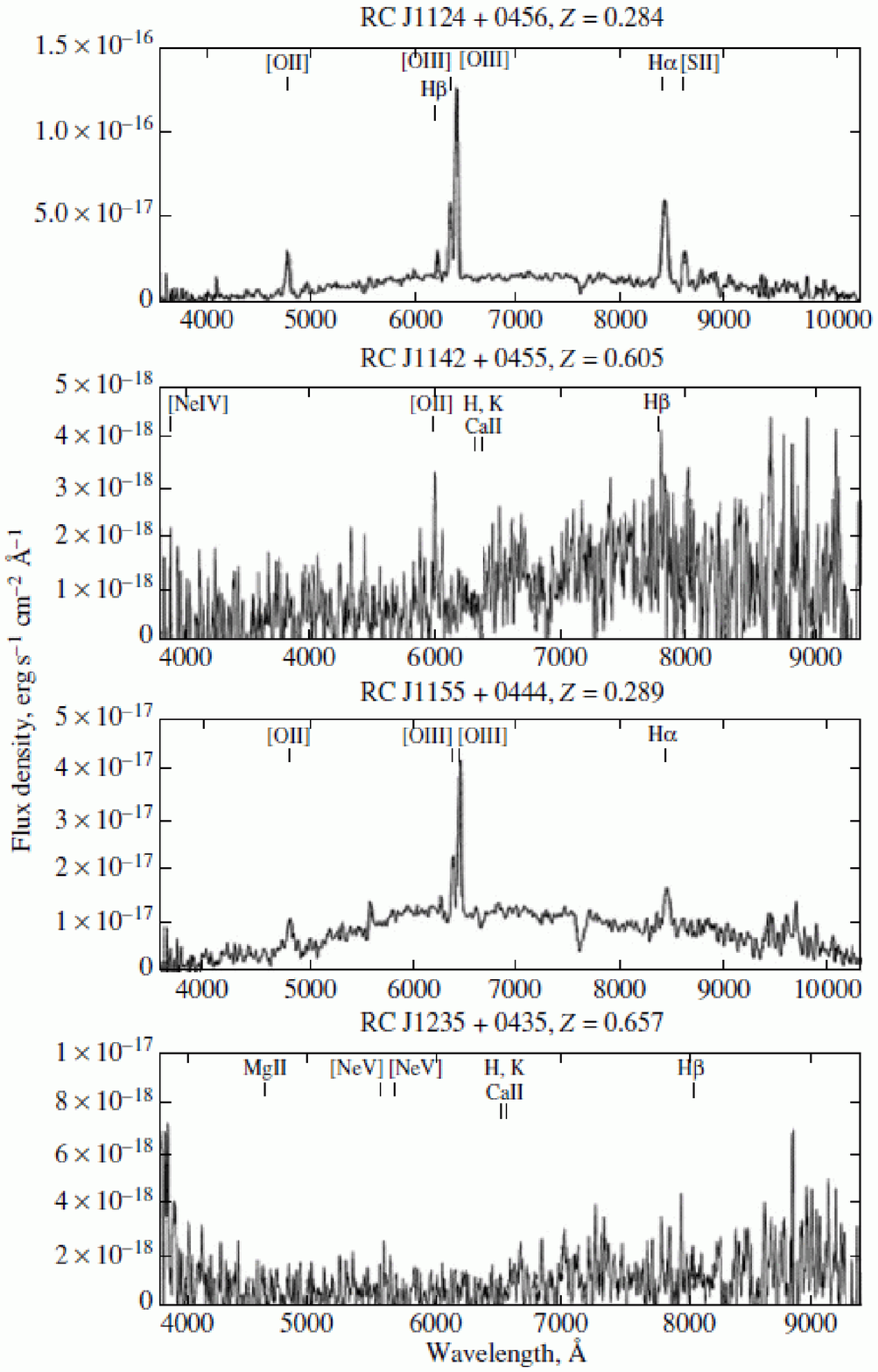,width=17cm}}
\centerline{{Fig. 2. (Continued)}}
\end{figure}
\begin{figure}
\centerline{\psfig{figure=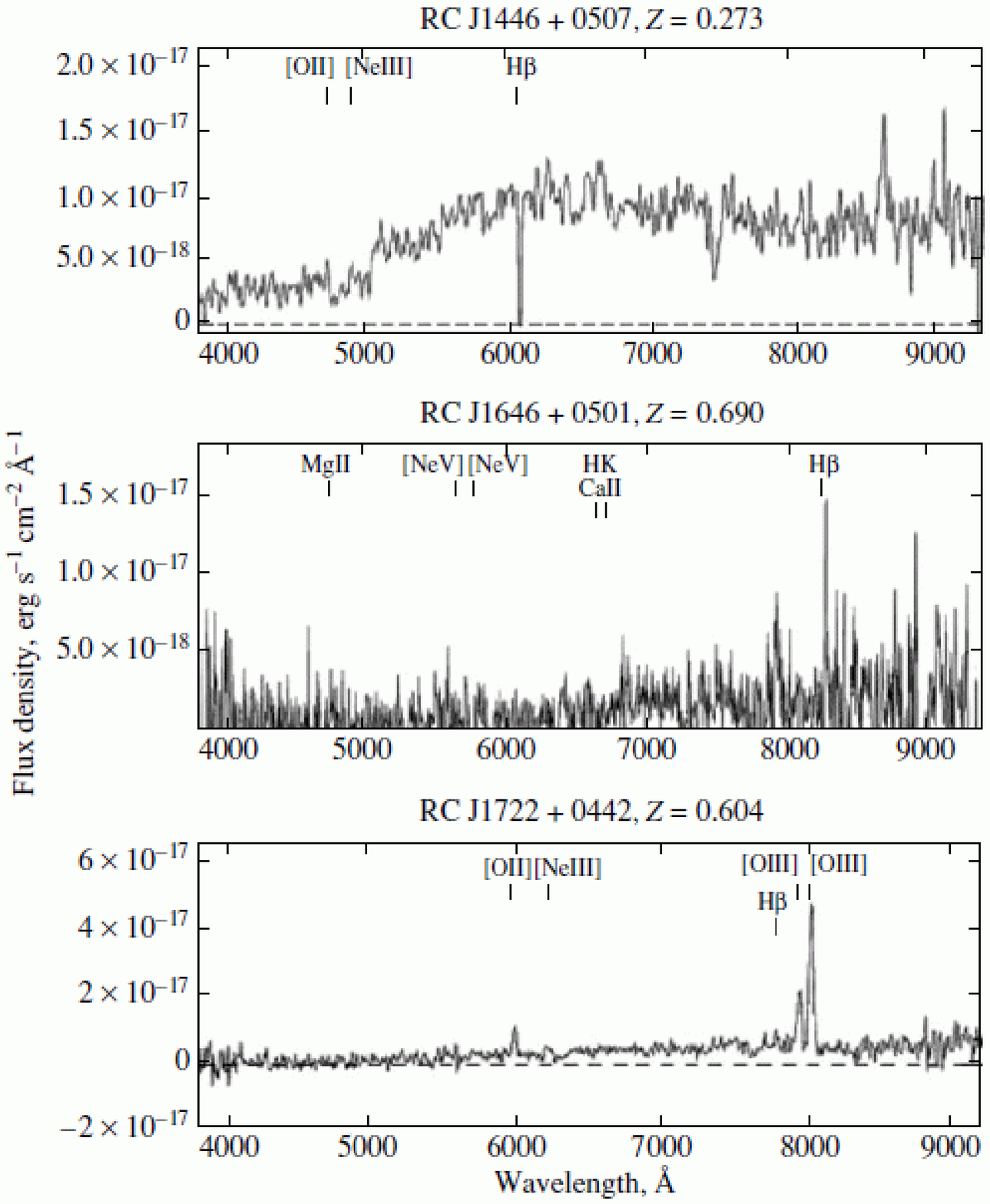,width=17cm}}
\centerline{{Fig. 2. (Continued)}}
\end{figure}

{\bf RCJ\,1011+0502} (Table 2). This is the faintest
galaxy among the RC objects, with the lowest luminosity
 in the optical (possibly a Sy1 galaxy with a 
strong jet). 

{\bf RCJ\,1051+0449} (Table 4). The direct images give
B=23.77, V=23.12, R=22.74, while the object
was not detected in the I-band (the images were taken
in the presence of seeing of 1.5''); no images of a
BVRI standard were taken, and the reduction was
made using surrounding SDSS objects. 

{\bf RCJ\,1100+0444} (Table 3). The NED-based redshift
is $z_{sp}$=0.8861.

{\bf RCJ\,1124+0456} (Table 2). The NED-based redshift
 is $z_{sp}$=0.2827.

{\bf RCJ\,1148+0455} (Table 4). The NED redshift,
$z_{sp}$=0.42,is not confirmed.

{\bf RCJ\,1152+0442} (Table 4). The photometric redshift
$z_{ph}$=1.24, disagrees with our results.

{\bf RCJ\,1333+0451} (Table 3). The NED-based redshift
is $z_{sp}$=1.4024; $z_{ph}$=1.07.

{\bf RCJ\,1503+0456} (Table 1). It is more likely that
the broad emission line at 5002\,\aa is MgII 2798\,\aa
than CIII] 1909\,\aa. The MgII identification is favored,
with a possibly detected continuum jump redward of 
7150\,\aa ($z_{sp}$=0.788). This is a BLRG, consistent
with its magnitude and radio morphology (core+jet);
$z_{ph}$=0.88.

{\bf RCJ\,1735+0454} (Table 4). No continuum or
bright emission lines were detected. If the weak 
emission at 7274\,\aa is [OII] 3727\,\aa, then the redshift
is $z_{sp}$=0.952. However, the spectrum to the red is
too short to see the continuum jump and confirm this 
hypothesis. 

{\bf RCJ\,1740+0502} (Table 3). This object has a
broad Ly$\alpha$ line (5560\,\aa) but no CIV line. This is a
star-like object in the optical, i.e., with a dominant 
nucleus. However, it is not a typical quasar; such 
objects are sometimes called WQs (weak quasars). 

{\bf RCJ\,2219+0458} (Table 4). The continuum grows
to the blue, in contradiction with the galaxy's colors
(B=24.8, V=25.03, R=23.72, I=22.25); $z_{ph}=1.24$.

{\bf RCJ\,2225+0523} (Table 3). The NED-based redshift
 is $z_{sp}=2.323$.

{\bf RCJ\,2247+0507} (Table 1). The spectrum contains a single broad
emission line, MgII 2798\,\aa.

The optical spectra of radio galaxies with $z>0.7$
and $z<0.7$ are shown in Figs.\,1 and 2. Figure 3
displays the spectra of the quasars. 

Thus, among the 22 radio galaxies in Table\,1,
we detected one object with $z=4.51$, one with $3\le z<4$
$z<4$,three with $2\le z<3$,
and four with $1\le z<2$.
Three more objects have redshifts close to unity
($>$0.9). The redshifts of 10 sources are in the range
from 0.74 to 0.87. The redshifts of the 15 radio galaxies
 from Table 2 are between 0.27 and 0.69. 
Note that there are no objects with LAS$>$12''
among the nine radio galaxies with $z>1$.The
500-MHz spectral indices of the objects with $z>1$
are the same as or lower than their spectral indices 
at 3940\,MHz. As a rule, the spectral indices of
such objects at 3940\,MHz exceed unity. Two of
the three objects with $z>3$ (RCJ\,0105+0501 and
RCJ\,1740+0502) have CL (core and lobe) structure,
while the third object (RCJ\,0311+0507) is a strongly
asymmetric triple, almost a CL source. 

One object among the 17 quasars has z=3.57
(RCJ\,1740+0502), four have $2\le z<3$,
seven have
$1\le z<2$,
four have $0.89Â\le z<1$,
and onehas z=0.72.

\begin{figure}
\centerline{\psfig{figure=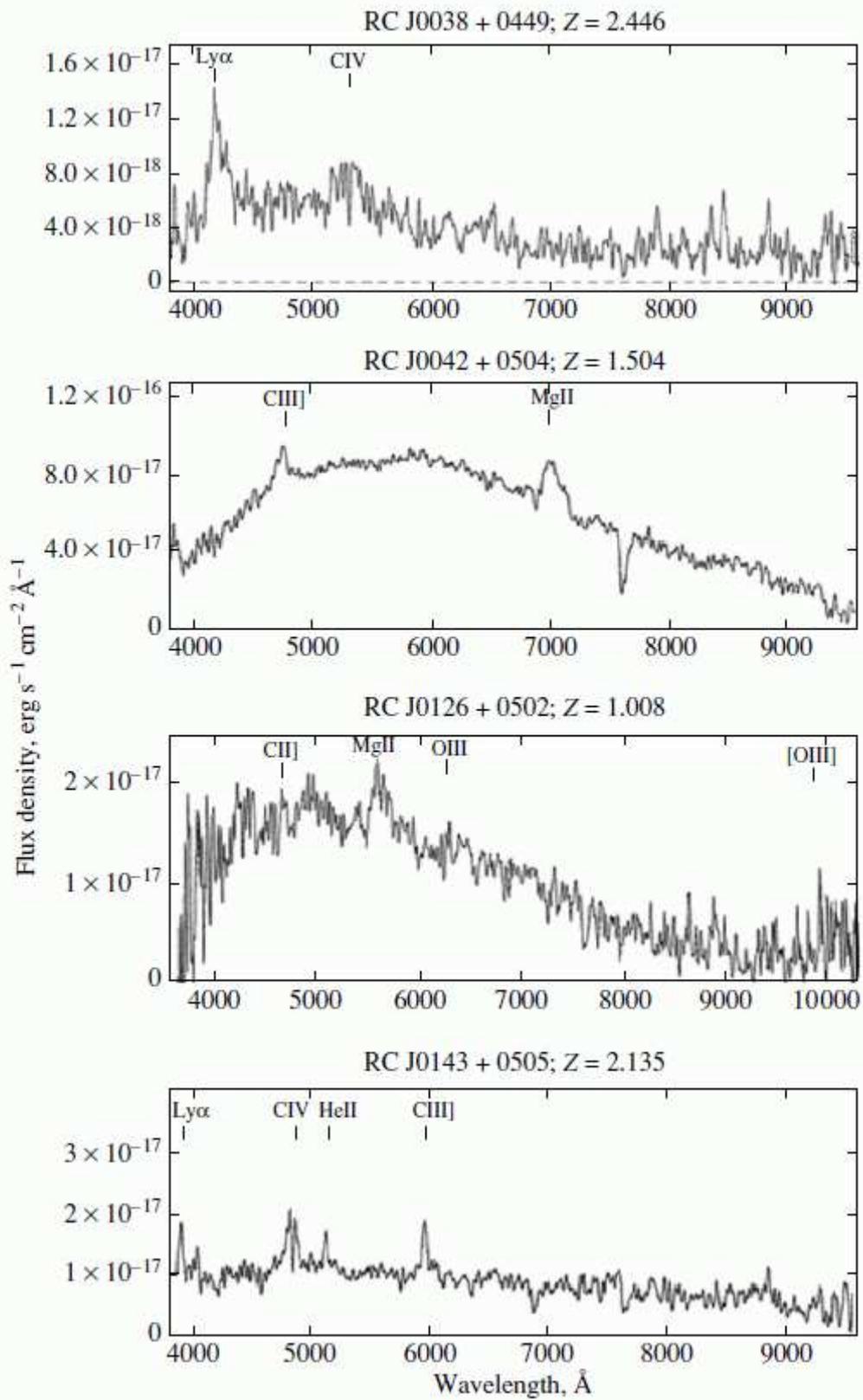,width=17cm}}
\caption{Spectra of quasars.}
\end{figure}
\begin{figure}
\centerline{{Fig. 3. (Continued)}}
\centerline{\psfig{figure=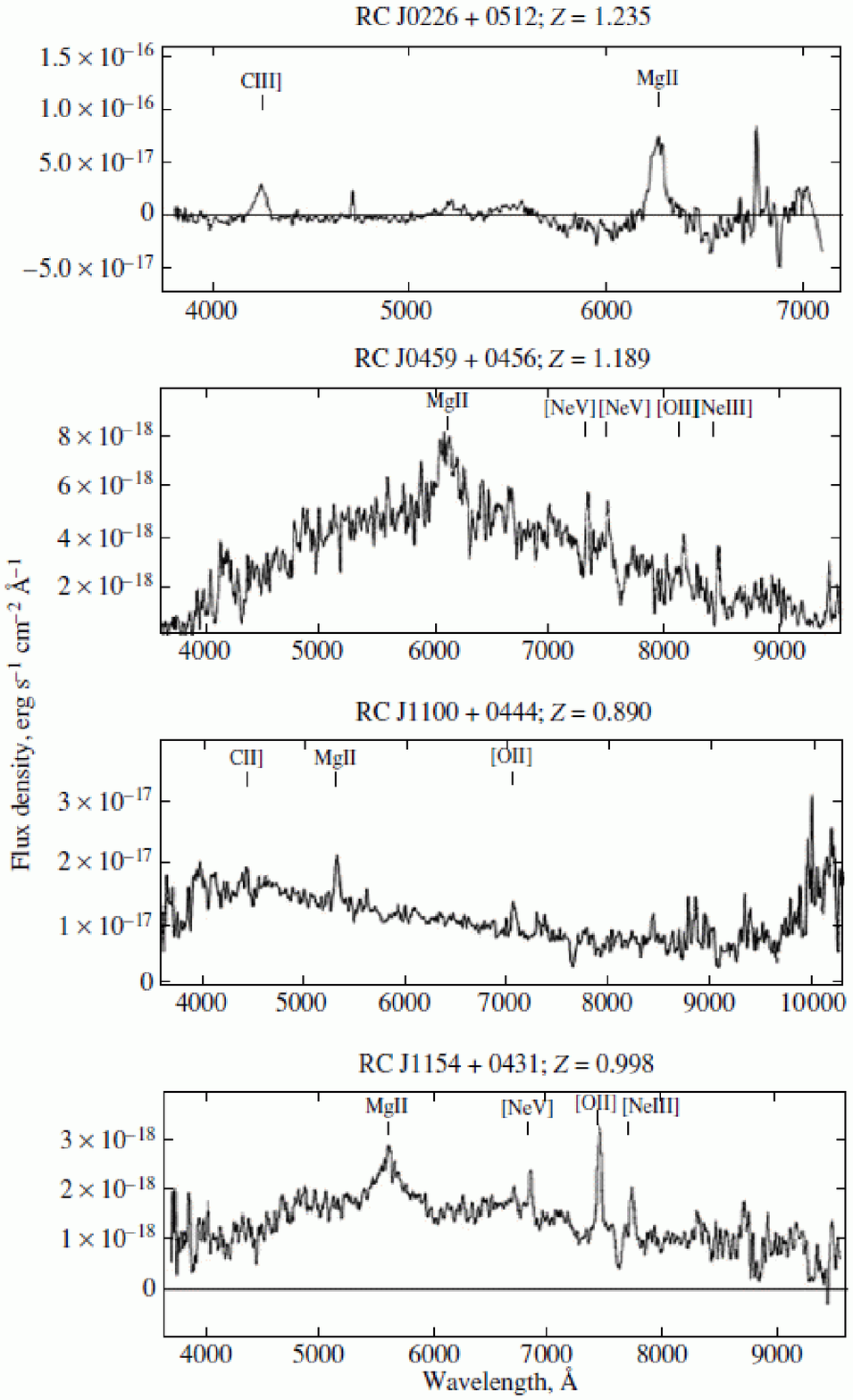,width=17cm}}
\end{figure}
\begin{figure}
\centerline{\psfig{figure=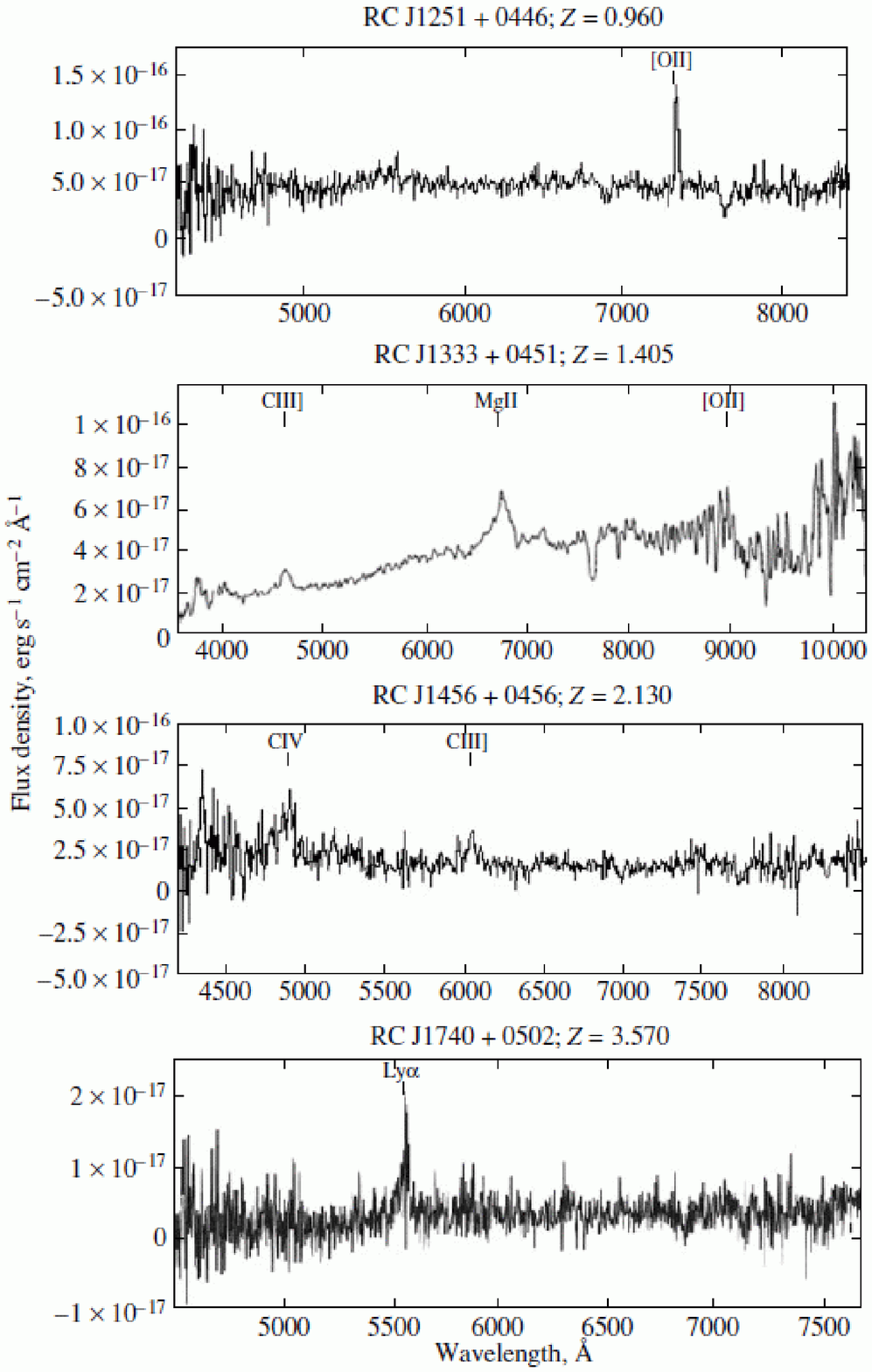,width=17cm}}
\centerline{{Fig. 3. (Continued)}}
\end{figure}
\begin{figure}
\centerline{\psfig{figure=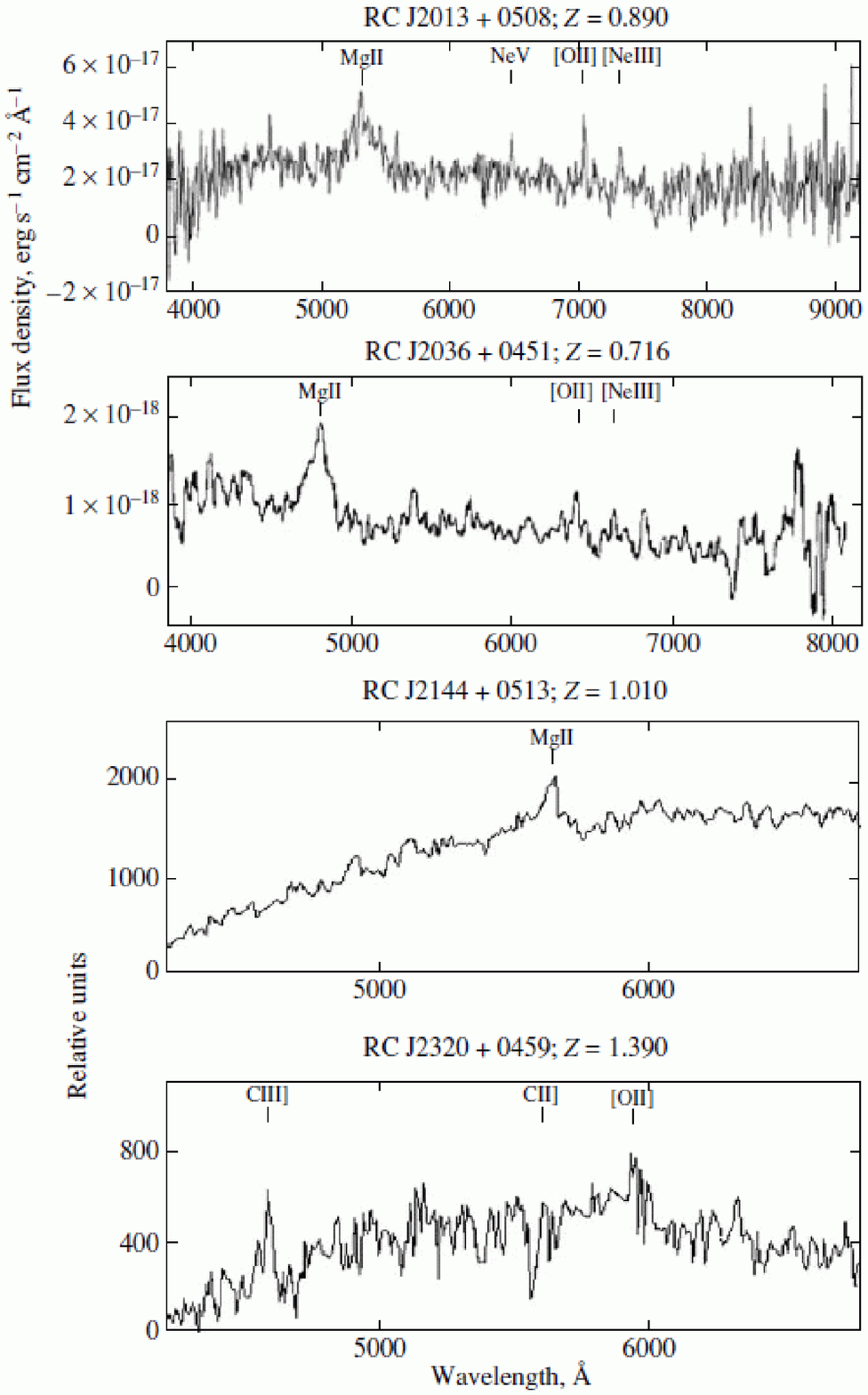,width=17cm}}
\centerline{{Fig. 3. (Continued)}}
\end{figure}

Figures 4 and 5 display the redshift distributions of
the radio galaxies and quasars.
 
\section*{3. CONCLUSIONS}

The ``Big Trio'' project includes about 100 objects,
for 71 of which we have obtained optical spectra
with the ``Scorpio'' spectrograph mounted on the 6-m
telescope of the SAO (17 quasars and 54 radio galaxies).

Of the studied radio galaxies, four have redshifts 
$1\le z<2$,
three have $2\le z<3$,
one has $3\le z<4$,
and one has z=4.51. Thirteen sources have $0.7<z<1$ and 15 have
$0. <z<0.7$. Five of the program
quasars have $0.7<z<1$, seven have $1\le z<2$,
four have $2< z<3$,and one has $z=3.57$. We detected
no spectral lines for 17 objects. 

Among the studied steep-and ultra-steep spectrum radio sources
with measured redshifts $z$
(54 objects), we found $\sim$39
to have $z>2$, $\sim$6\% to have $z>3$, and $\sim$2\% to have
$z>4.5$ (4.514). The total number of radio galaxies
with redshifts $z>4$ detected to date is currently
six [28].

The failure to detect spectral lines in the spectra of
17 radio galaxies may indicate that they have redshifts
in the range of $1.5<z<2$ or $z>5$, or may be due
to absorption by dust. In some cases, observational
errors or poor weather conditions could also be responsible.

Our data confirm the effectiveness of identifying
candidate distant objects based on the properties of
their radio and optical continua (steep radio spectra,
characteristics of FRII objects, small angular sizes,
large ratios of the radio and optical luminosities).
These criteria made it possible to detect the
object RCJ\,0311+0507 with $z=4.514$, which has a
very high radio luminosity at centimeter wavelengths. 
Note that, since there are no selection effects, the 
fraction of distant objects in modern radio surveys 
(such as the NVSS) is much lower than the fraction 
in Fig.\,4 resulting from the ``Big Trio'' program (see
also the review by Pedani [29]). The identification of 
FRII radio sources at high redshifts is also helpful 
for searches for early ``giant black holes'' or
first-generation clusters of galaxies.

We have presented here factual information on the 
71 studied ``Big Trio'' objects. We are planning to
discuss astrophysical implications of these data in 
future publications. 

\begin{figure}
\centerline{\psfig{figure=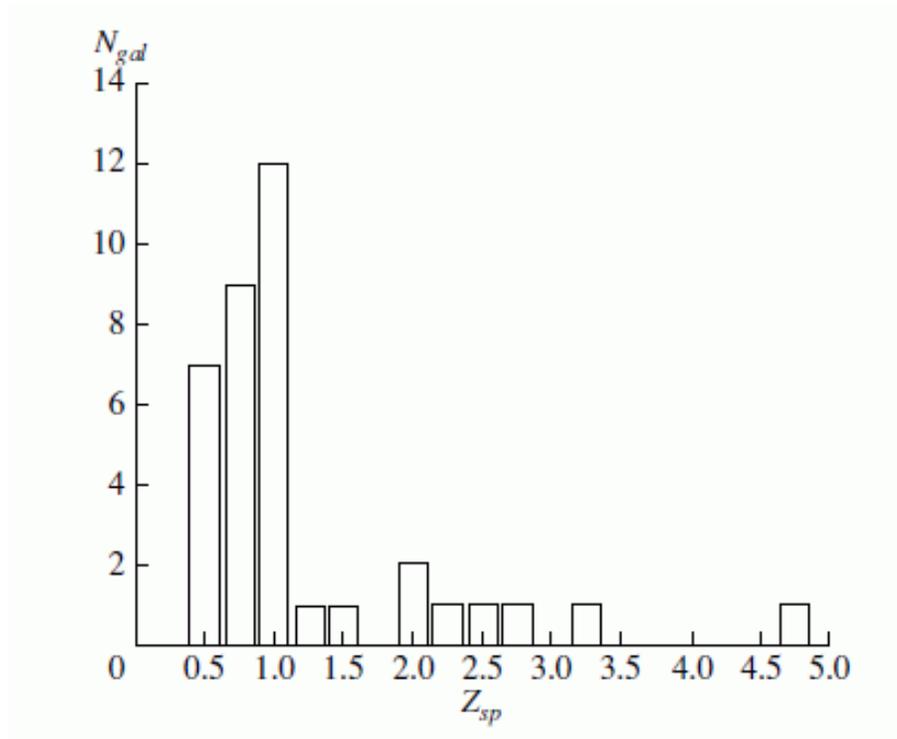,width=12cm}}
\caption{Redshift distribution of the galaxies.}
\end{figure}

\begin{figure}
\centerline{\psfig{figure=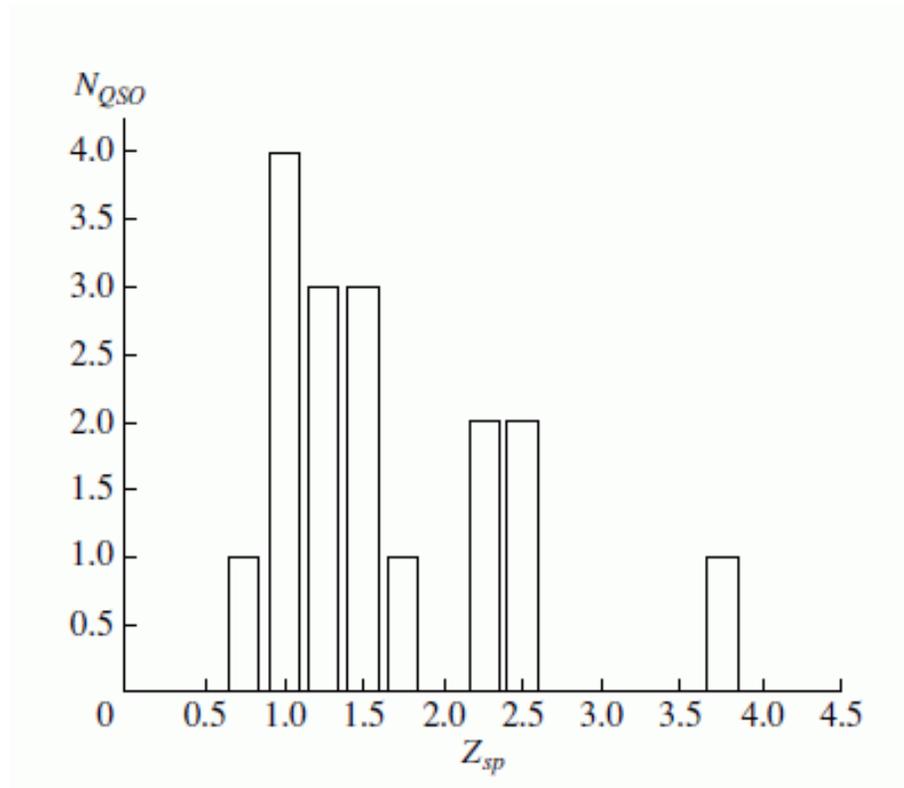,width=12cm}}
\caption{Redshift distribution of the quasars.}
\end{figure}

\vspace*{2ex}
{\bf Acknowledgments}

This study was partially supported by the Russian
 Foundation for Basic Research (project nos. 08-02-00486a,
09-07-00320) and the Program of State
Support of Leading Scientific Schools of the Russian
 Federation. The authors thank the observers at 
the 6-m telescope of the SAO who performed the 
observations with the ``Scorpio'' spectrograph. The
Very Large Array of the National Radio Astronomy 
Observatory is a Facility of the National Science 
Foundation operated under cooperative agreement by 
Associated Universities, Inc., a science management 
corporation.

{\it Translated by N. Samus.}

\end{document}